\documentclass[conference]{IEEEtran}
\IEEEoverridecommandlockouts

\usepackage{cite}
\usepackage{amsmath,amssymb,amsfonts}
\usepackage{algorithmic}
\usepackage{graphicx}
\usepackage{threeparttable}
\usepackage{textcomp}
\usepackage{xcolor}
\def\BibTeX{{\rm B\kern-.05em{\sc i\kern-.025em b}\kern-.08em
    T\kern-.1667em\lower.7ex\hbox{E}\kern-.125emX}}
\begin{document}



\title{EEG Spectral Analysis in Gray Zone Between Healthy and Insomnia\\
\thanks{This work was partly supported by Institute of Information \& Communications Technology Planning \& Evaluation (IITP) grant funded by the Korea government (MSIT) (No. RS--2021--II--212068, Artificial Intelligence Innovation Hub, No. RS--2024--00336673, AI Technology for Interactive Communication of Language Impaired Individuals, and No. RS--2019--II190079, Artificial Intelligence Graduate School Program (Korea University)).}
}


\author{

\IEEEauthorblockN{Ha-Na Jo}
\IEEEauthorblockA{\textit{Dept. of Artificial Intelligence} \\
\textit{Korea University} \\
Seoul, Republic of Korea \\ 
hn\_jo@korea.ac.kr}

\and

\IEEEauthorblockN{Young-Seok Kweon}
\IEEEauthorblockA{\textit{Dept. of Brain and Cognitive Engineering} \\
\textit{Korea University} \\
Seoul, Republic of Korea \\
youngseokkweon@korea.ac.kr} \\

\and

\IEEEauthorblockN{Seo-Hyun Lee}
\IEEEauthorblockA{\textit{Dept. of Brain and Cognitive Engineering} \\
\textit{Korea University} \\
Seoul, Republic of Korea \\
seohyunlee@korea.ac.kr} \\
}

\maketitle

\begin{abstract}
This study investigates the sleep characteristics and brain activity of individuals in the gray zone of insomnia, a population that experiences sleep disturbances yet does not fully meet the clinical criteria for chronic insomnia. Thirteen healthy participants and thirteen individuals from the gray zone were assessed using polysomnography and electroencephalogram to analyze both sleep architecture and neural activity. Although no significant differences in objective sleep quality or structure were found between the groups, gray zone individuals reported higher insomnia severity index scores, indicating subjective sleep difficulties. Electroencephalogram analysis revealed increased delta and alpha activity during the wake stage, suggesting lingering sleep inertia, while non-rapid eye movement stages 1 and 2 exhibited elevated beta and gamma activity, often associated with chronic insomnia. However, these high-frequency patterns were not observed in non-rapid eye movement stage 3 or rapid eye movement sleep, suggesting less severe disruptions compared to chronic insomnia. This study emphasizes that despite normal polysomnography findings, EEG patterns in gray zone individuals suggest a potential risk for chronic insomnia, highlighting the need for early identification and tailored intervention strategies to prevent progression.
\end{abstract}


\section{INTRODUCTION}
Insomnia is a prevalent sleep disorder that can significantly impair various aspects of an individual's life. The condition has been associated with a range of detrimental effects, including difficulties in emotional regulation, decreased cognitive performance, and increased risk of mood disorders such as anxiety and depression \cite{intro1}. Research indicates that individuals suffering from insomnia often experience heightened emotional reactivity and impaired coping mechanisms, leading to a diminished quality of life. Given the growing prevalence of insomnia and its associated repercussions, understanding the underlying mechanisms of this disorder is of paramount importance. By elucidating the emotional and physiological characteristics of insomnia, researchers can better address its impact and develop effective interventions for those affected.

In recent years, research has increasingly focused on the relationship between sleep disorders like insomnia and brain function. Given the central role the brain plays in regulating sleep, brain-computer interface (BCI) technologies have emerged as a promising area of study in understanding and treating insomnia \cite{prml60-1, intro2}. BCI research investigates how neural activity correlates with sleep disturbances, aiming to create systems that can monitor and potentially modulate brain states to improve sleep quality \cite{prml60-2}. This approach is particularly relevant for identifying biomarkers that distinguish different types of insomnia and could lead to personalized interventions.

Despite the advancements in understanding insomnia, the majority of existing research has predominantly focused on individuals with chronic insomnia. This narrow scope presents several limitations, particularly the significant demographic differences between control and patient groups, which may confound the interpretation of findings \cite{intro3}. Furthermore, there is a notable paucity of studies investigating individuals who experience insomnia-like symptoms yet do not meet the criteria for chronic insomnia. These individuals may endure disturbances in sleep patterns and daytime functioning but are often overlooked in both clinical and research contexts.

In light of these gaps in the literature, the present study aims to explore the characteristics of a gray zone group—individuals who do not qualify as chronic insomnia patients but report experiencing discomfort in daily functioning due to sleep disturbances. This gray zone group may represent a transitional state, placing individuals at increased risk of developing chronic insomnia. By analyzing this unique cohort, the current research seeks to identify specific features that distinguish them from both healthy individuals and those with chronic insomnia. Ultimately, this study aspires to lay the groundwork for targeted interventions and care strategies for this at-risk population, thereby potentially mitigating the progression to chronic insomnia and enhancing overall sleep health.

\begin{figure*}[thpb]
  \centering
  \includegraphics[width=\textwidth]{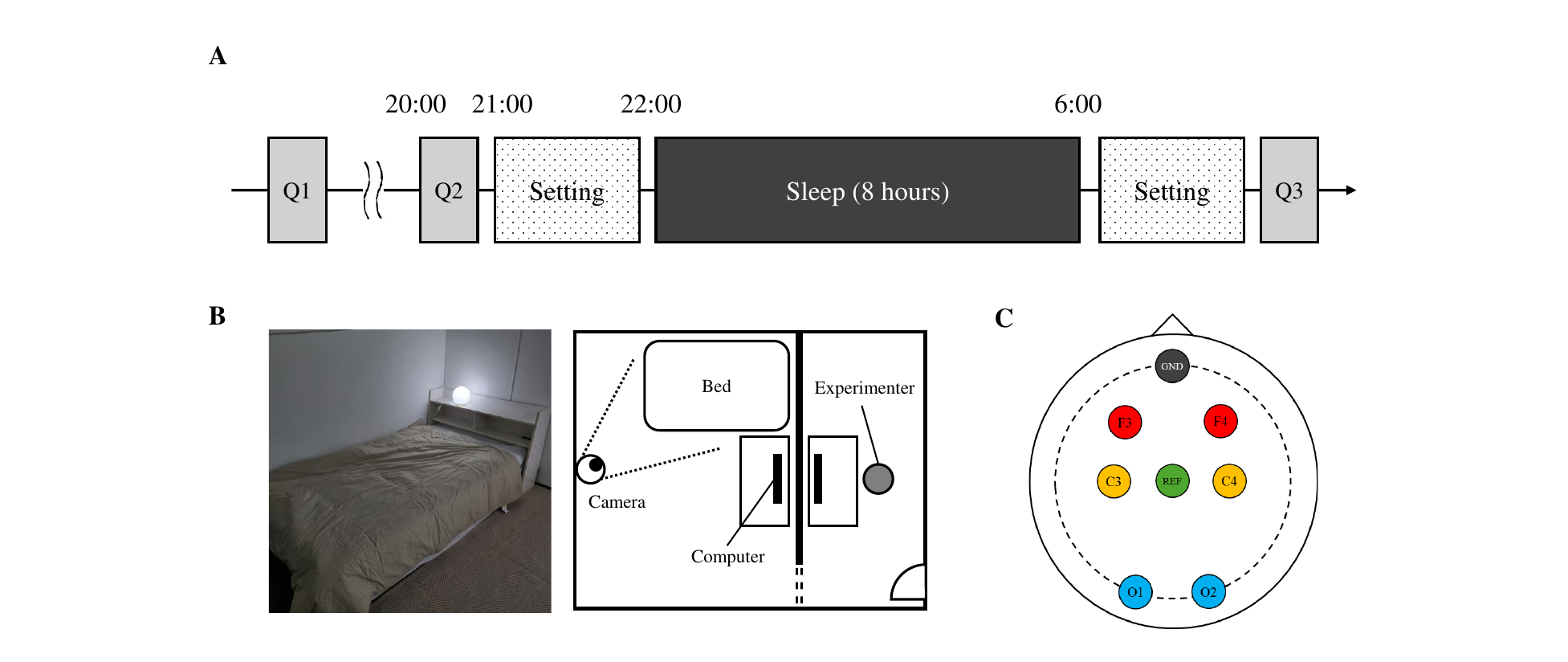}
  \caption{Experimental setting. a) Experimental procedure. Q=questionnaire, b) Experimental environment. c) Segmentation of brain regions. The brain region is divided into three groups: F=frontal, C=central, and O=occipital.}
  \label{figure1}
\end{figure*}

\section{METHODS}
\subsection{Experimental Procedure}
The study involved 26 male participants with a mean age of 27.9 years (standard deviation $\pm$ 2.67 years). All participants were healthy individuals without any history of insomnia or other sleep disorders. Prior to the start of the experiment, ethical approval was granted by the Institutional Review Board at Korea University (KUIRB-2022-0222-01), and each participant provided written informed consent, following previous study \cite{prml60-3}.

Participants were instructed to complete a basic questionnaire the day before the experiment. On the day of the experiment, they arrived at the laboratory at 6:00 PM, changed into comfortable clothing, and received a briefing on the experimental procedure and precautions, allowing time to acclimate to the laboratory environment. At 8:00 PM, a preliminary questionnaire was conducted, followed by the attachment of electrodes. The sleep session began at 10:00 PM and lasted until 6:00 AM the following day, providing a total of 8 hours of sleep. 

The laboratory was divided into two spaces: one for sleeping and one for real-time monitoring of sleep. To address any issues that might arise during the experiment, two personnel—a sleep expert and the experiment supervisor—remained present in the laboratory throughout the session.  To control for confounding variables, participants were required to maintain a minimum of 7 hours of sleep per night for one week prior to the experiment \cite{method1}. Additionally, alcohol and caffeine consumption were prohibited from the day before the experiment through to the day of the experiment. After the experiment concluded, participants completed a brief questionnaire.

\begin{table}[]
\centering
\caption{Comparison of sleep architectures between healthy group and gray zone group.}
\label{table1}
\resizebox{\columnwidth}{!}{%
\begin{tabular}{cccc}
\hline
\textbf{Component}    & \textbf{\begin{tabular}[c]{@{}c@{}}Healthy\\ (\textit{n}=13)\end{tabular}} & \textbf{\begin{tabular}[c]{@{}c@{}}Gray zone\\ (\textit{n}=13)\end{tabular}} & \textbf{\textit{p}-value} \\ \hline
TRT (min.)            & 489.34                                                            & 391.00                                                              & 0.43             \\
TST (min.)            & 373.22                                                            & 379.00                                                              & 0.84             \\
Sleep efficiency (\%) & 77.31                                                             & 77.93                                                               & 0.90             \\
SOL (min.)            & 7.40                                                              & 12.85                                                               & 0.28             \\
WASO (min.)           & 102.48                                                            & 94.19                                                               & 0.73             \\
NREM 1 (\% of TST)     & 21.81                                                             & 23.56                                                               & 0.60             \\
NREM 2 (\% of TST)     & 58.00                                                             & 55.79                                                               & 0.45             \\
NREM 3 (\% of TST)     & 1.70                                                              & 2.13                                                                & 0.56             \\
REM (\% of TST)       & 18.49                                                             & 18.50                                                               & 0.99             \\
AHI                   & 7.41                                                              & 12.24                                                               & 0.29             \\
PLMD                  & 6.96                                                              & 11.49                                                               & 0.28             \\ \hline
\end{tabular}%
}
\begin{tablenotes}[flushleft]
\item TRT=total recording time, TST=total sleep time, SOL=sleep onset latency, WASO=wake after sleep onset, NREM=non-rapid eye movement stage, REM=rapid eye movement sleep, AHI=apnea-hypopnea index, PLMD=periodic limb movement disorder
\end{tablenotes}
\end{table}

\begin{figure*}[thpb]
  \centering
  \includegraphics[width=\textwidth]{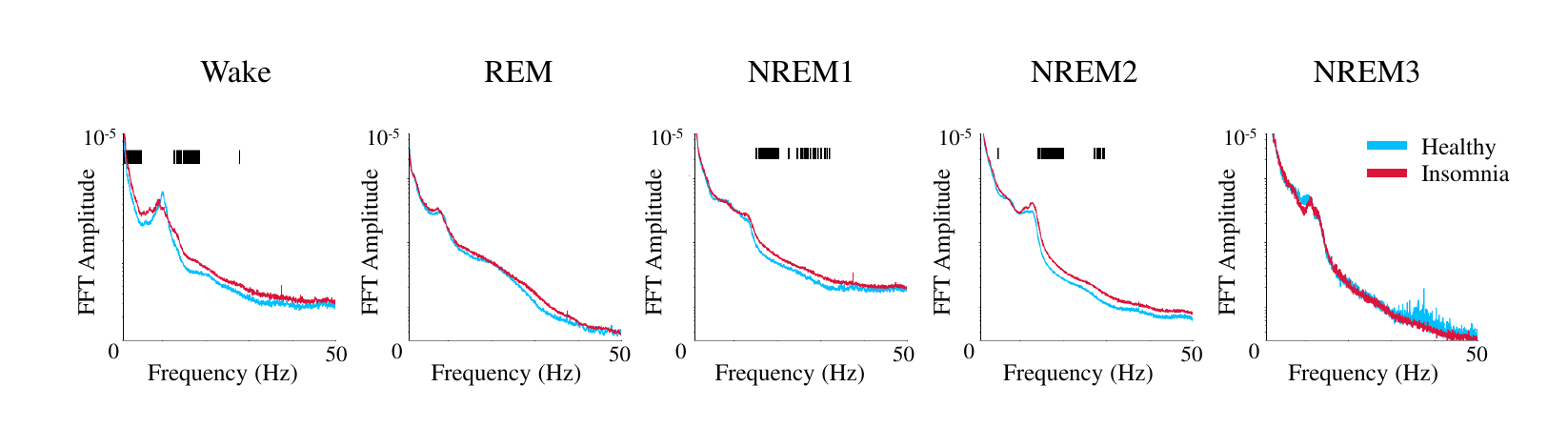}
  \caption{Result of power spectral analysis. Black box represents the statistical significance (\textit{p}$<$0.05). FFT=fast Fourier transform, REM=rapid eye movement sleep, NREM=non-rapid eye movement stage. }
  \label{figure2}
\end{figure*}

\subsection{Polysomnography}
Both the control and gray zone groups underwent polysomnography (PSG) as part of this study. PSG involved the use of multiple electrodes and sensors to record a wide range of physiological signals. These included an electroencephalogram (EEG) to monitor brain activity, electrooculograms to track eye movements, an electrocardiogram to measure heart activity, and a chin electromyogram to detect muscle tone. Additionally, oximetry was used to assess blood oxygen levels, inductance plethysmography was employed to monitor chest wall and abdominal movements, and airflow was measured from both the nose and mouth \cite{prml1}.

The sleep stages were identified and classified based on the EEG and other physiological data. Following international standards set by the American Academy of Sleep Medicine (AASM), sleep experts manually labeled 30-second epochs of the PSG recordings. The stages were categorized into Wake, REM (rapid eye movement), and NREM (non-rapid eye movement) stages 1, 2, and 3. NREM 1 (N1) represents light sleep, NREM 2 (N2) is a deeper stage of sleep, and NREM 3 (N3) refers to slow-wave, deep sleep \cite{method2}.

From these labeled recordings, objective sleep parameters were extracted, including total sleep time (TST), wake after sleep onset (WASO), sleep efficiency (SE), and sleep onset latency (SOL). TST was calculated as the total time spent asleep, excluding any periods of wakefulness during the night. WASO measured the amount of time spent awake after the initial onset of sleep. SE was determined as the percentage of time spent asleep out of the total recording time. SOL was defined as the time from lights off to the first epoch classified as N1 or higher. The duration of each sleep stage was also recorded based on the time spent in each respective stage.

\subsection{Statistical Analysis}
To investigate the spatial characteristics of EEG data, we divided the brain into three specific regions: frontal, central, and occipital regions. We then conducted our analysis for each of these designated areas. The use of power spectral density (PSD) is crucial as it enables us to thoroughly analyze and quantify the frequency distribution and power levels present in a signal. This is vital for gaining insights into the underlying dynamics of various neural and physiological processes. Additionally, prior to this analysis, the EEG signals underwent preprocessing steps, including noise reduction and filtering to isolate the relevant frequency bands \cite{prml60-4, prml60-5, result1}.

The paired \textit{t}-test is a statistical technique for comparing two related groups to determine if there is a significant difference between them, such as measurements taken from the same subjects under varying conditions or time points. We used this test to analyze the statistical significance of differences across groups, establishing a significance threshold at \textit{p}$<$0.05 \cite{prml60-6}.

\section{RESULTS AND DISCUSSION}
\subsection{Sleep Architecture}
A comparative analysis of the demographic characteristics revealed that the mean ages of the participants in the healthy group and the gray zone group were 28.6 years and 27.3 years, respectively. The difference in age between the two groups was not statistically significant (\textit{p}=0.23). When examining the sleep architecture, several parameters were assessed. While differences in the average sleep structure were noted between the groups, these differences did not reach statistical significance (\textit{p}$>$0.05). This indicates that although variations in sleep architecture exist, they are not substantial enough to be considered clinically relevant. Furthermore, the analysis of sleep quality showed no statistically significant differences between the two groups (\textit{p}$>$0.05).

Contrary to most previous studies, there were no statistically significant differences in the demographic variables such as gender or age between the healthy control group and the gray zone group \cite{result1}. This outcome allows us to exclude any neurophysiological variances potentially driven by gender or age, making the comparison between these groups more robust.

The Insomnia Severity Index (ISI) score, a subjective measure of insomnia severity, revealed a significant difference between the two groups. Healthy participants had an average ISI score of 4.53, while the gray zone group scored 9.53, which was statistically significant (\textit{p}$<$0.05). This suggests that individuals in the gray zone experience subjective symptoms of insomnia. However, objective measures of sleep—such as TST, stage-specific sleep times, and sleep quality scores—did not reflect significant differences between the two groups (\textit{p}$>$0.05).

This discrepancy between subjective insomnia symptoms and objective sleep measures suggests that individuals in the gray zone, while feeling the effects of insomnia, are not exhibiting the traditional sleep disruptions detectable via PSG or other basic sleep architecture analyses \cite{result2}. This highlights the gray zone group's position between healthy sleepers and chronic insomnia patients, demonstrating that subjective sleep experiences may not always align with objective findings in early-stage insomnia development.

\subsection{Brain Activation by Sleep Stage}
All results presented herein are based on the average values obtained from all EEG channels. During the wake stage, the analysis revealed that gray zone participants exhibited statistically significant increases in delta and alpha band activity compared to healthy individuals (\textit{p}$<$0.05) \cite{prml2} . This heightened delta activity during wakefulness suggests a phenomenon known as sleep inertia, which is characterized by difficulty transitioning from sleep to wakefulness \cite{result3, result7, result4}. As a result, individuals in the gray zone may experience some symptoms associated with insomnia, such as prolonged sleep onset and impaired cognitive function upon waking.

Gray zone individuals showed significantly higher beta and gamma band activation in both NREM1 and NREM2 stages compared to healthy participants. While low-frequency delta and theta waves typically dominate NREM stages, this increased high-frequency activity suggests potential disruption in sleep architecture, similar to chronic insomnia. However, unlike chronic insomnia patients, gray zone individuals had fewer channels with high-frequency activation, and no significant activity was observed in the N3 stage, indicating a milder degree of disruption \cite{result5, result6}.

Interestingly, while the EEG patterns of the gray zone group exhibit non-specific brain wave activity, the lower intensity of these patterns may not sufficiently alter their overall sleep structure or quality. Nonetheless, the similarities in EEG characteristics between the gray zone group and chronic insomnia group suggest that this group may represent a risk population for developing chronic sleep disorders.

\section{CONCLUSION}

Our study underscores the importance of recognizing the gray zone of insomnia as a distinct category, marked by unique EEG patterns that differentiate it from both healthy sleep and chronic insomnia. The lack of significant differences in traditional sleep architecture reveals the limitations of relying solely on PSG for detecting critical neurological activity. By refining our understanding of the gray zone, we can better identify those at risk of chronic insomnia and lay the groundwork for targeted interventions. Machine learning models could improve precision in identifying these individuals through EEG data and other physiological markers \cite{prml3, prml60-7}, enabling timely interventions and allowing clinicians to effectively monitor sleep disorder progression. Future studies should focus on longitudinal tracking of individuals in the gray zone to assess early intervention efficacy.



\bibliographystyle{IEEEtran}
\bibliography{ref}

\begin{thebibliography}{10}
\providecommand{\url}[1]{#1}
\csname url@samestyle\endcsname
\providecommand{\newblock}{\relax}
\providecommand{\bibinfo}[2]{#2}
\providecommand{\BIBentrySTDinterwordspacing}{\spaceskip=0pt\relax}
\providecommand{\BIBentryALTinterwordstretchfactor}{4}
\providecommand{\BIBentryALTinterwordspacing}{\spaceskip=\fontdimen2\font plus
\BIBentryALTinterwordstretchfactor\fontdimen3\font minus \fontdimen4\font\relax}
\providecommand{\BIBforeignlanguage}[2]{{%
\expandafter\ifx\csname l@#1\endcsname\relax
\typeout{** WARNING: IEEEtran.bst: No hyphenation pattern has been}%
\typeout{** loaded for the language `#1'. Using the pattern for}%
\typeout{** the default language instead.}%
\else
\language=\csname l@#1\endcsname
\fi
#2}}
\providecommand{\BIBdecl}{\relax}
\BIBdecl

\bibitem{intro1}
M.~Younes, A.~Azarbarzin, M.~Reid, D.~R. Mazzotti, and S.~Redline, ``Characteristics and reproducibility of novel sleep {EEG} biomarkers and their variation with sleep apnea and insomnia in a large community-based cohort,'' \emph{Sleep}, vol.~44, no.~10, p. zsab145, 2021.

\bibitem{prml60-1}
D.-H. Lee, J.-H. Jeong, K.~Kim, B.-W. Yu, and S.-W. Lee, ``Continuous {EEG} decoding of pilots’ mental states using multiple feature block-based convolutional neural network,'' \emph{IEEE Access}, vol.~8, pp. 121\,929--121\,941, 2020.

\bibitem{intro2}
M.~Sharma, V.~Patel, and U.~R. Acharya, ``Automated identification of insomnia using optimal bi-orthogonal wavelet transform technique with single-channel {EEG} signals,'' \emph{Knowl.-Based Syst.}, vol. 224, p. 107078, 2021.

\bibitem{prml60-2}
S.~K. Prabhakar, H.~Rajaguru, and S.-W. Lee, ``A framework for schizophrenia {EEG} signal classification with nature inspired optimization algorithms,'' \emph{IEEE Access}, vol.~8, pp. 39\,875--39\,897, 2020.

\bibitem{intro3}
Y.~Guo \emph{et~al.}, ``Thalamic network under wakefulness after sleep onset and its coupling with daytime fatigue in insomnia disorder: An {EEG-fMRI} study,'' \emph{J. Affect. Disord.}, vol. 334, pp. 92--99, 2023.

\bibitem{prml60-3}
J.-H. Jeong, B.-W. Yu, D.-H. Lee, and S.-W. Lee, ``Classification of drowsiness levels based on a deep spatio-temporal convolutional bidirectional {LSTM} network using electroencephalography signals,'' \emph{Brain Sci.}, vol.~9, no.~12, p. 348, 2019.

\bibitem{method1}
C.-H. Kao \emph{et~al.}, ``Insomnia subtypes characterised by objective sleep duration and {NREM} spectral power and the effect of acute sleep restriction: An exploratory analysis,'' \emph{Sci. Rep.}, vol.~11, no.~1, p. 24331, 2021.

\bibitem{prml1}
Y.-S. Kweon, G.-H. Shin, H.-G. Kwak, and H.-N. Jo, ``Multi-signal reconstruction using masked autoencoder from {EEG} during polysomnography,'' in \emph{Proc. IEEE Int. Winter Conf. Brain-Comput. Interface (BCI)}, 2024, pp. 1--4.

\bibitem{method2}
S.~F. Quan \emph{et~al.}, ``The sleep heart health study: Design, rationale, and methods,'' \emph{Sleep}, vol.~20, no.~12, pp. 1077--1085, 1997.

\bibitem{prml60-4}
S.-Y. Han, N.-S. Kwak, T.~Oh, and S.-W. Lee, ``Classification of pilots’ mental states using a multimodal deep learning network,'' \emph{Biocybern. Biomed. Eng.}, vol.~40, no.~1, pp. 324--336, 2020.

\bibitem{prml60-5}
R.~Mane, N.~Robinson, A.~P. Vinod, S.-W. Lee, and C.~Guan, ``A multi-view {CNN} with novel variance layer for motor imagery brain computer interface,'' in \emph{Proc. Int. Conf. IEEE Eng. Med. Biol. Soc. (EMBC)}, 2020, pp. 2950--2953.

\bibitem{result1}
W.~Kweon \emph{et~al.}, ``Amygdala resting-state functional connectivity alterations in patients with chronic insomnia disorder: Correlation with electroencephalography beta power during sleep,'' \emph{Sleep}, vol.~46, no.~10, p. zsad205, 2023.

\bibitem{prml60-6}
J.~Kim \emph{et~al.}, ``Abstract representations of associated emotions in the human brain,'' \emph{J. Neurosci.}, vol.~35, no.~14, pp. 5655--5663, 2015.

\bibitem{result2}
M.~Younes \emph{et~al.}, ``Sleep architecture based on sleep depth and propensity: Patterns in different demographics and sleep disorders and association with health outcomes,'' \emph{Sleep}, vol.~45, no.~6, p. zsac059, 2022.

\bibitem{prml2}
M.~Lee \emph{et~al.}, ``Connectivity differences between consciousness and unconsciousness in non-rapid eye movement sleep: A {TMS--EEG} study,'' \emph{Sci. Rep.}, vol.~9, no.~1, p. 5175, 2019.

\bibitem{result3}
R.~Vallat, D.~Meunier, A.~Nicolas, and P.~Ruby, ``Hard to wake up? the cerebral correlates of sleep inertia assessed using combined behavioral, {EEG and fMRI} measures,'' \emph{Neuroimage}, vol. 184, pp. 266--278, 2019.

\bibitem{result7}
M.~Navarrete \emph{et~al.}, ``Examining the optimal timing for closed-loop auditory stimulation of slow-wave sleep in young and older adults,'' \emph{Sleep}, vol.~43, no.~6, p. zsz315, 2020.

\bibitem{result4}
W.~Zhao \emph{et~al.}, ``{EEG} spectral analysis in insomnia disorder: A systematic review and meta-analysis,'' \emph{Sleep Med. Rev.}, vol.~59, p. 101457, 2021.

\bibitem{result5}
Y.~Shi \emph{et~al.}, ``Elevated beta activity in the nighttime sleep and multiple sleep latency electroencephalograms of chronic insomnia patients,'' \emph{Front. Neurosci.}, vol.~16, p. 1045934, 2022.

\bibitem{result6}
M.~L. Perlis, M.~T. Smith, P.~J. Andrews, H.~Orff, and D.~E. Giles, ``Beta/gamma {EEG} activity in patients with primary and secondary insomnia and good sleeper controls,'' \emph{Sleep}, vol.~24, no.~1, pp. 110--117, 2001.

\bibitem{prml3}
J.-H. Cho, J.-H. Jeong, K.-H. Shim, D.-J. Kim, and S.-W. Lee, ``Classification of hand motions within {EEG} signals for non-invasive {BCI}-based robot hand control,'' in \emph{Proc. IEEE Int. Conf. Syst. Man Cybern. (SMC)}, 2018, pp. 515--518.

\bibitem{prml60-7}
K.~Lee, S.-A. Kim, J.~Choi, and S.-W. Lee, ``Deep reinforcement learning in continuous action spaces: A case study in the game of simulated curling,'' in \emph{Proc. Int. Conf. Mach. Learn. (ICML)}, 2018, pp. 2937--2946.

\end{thebibliography}


\end{document}